\documentclass[a4paper]{jpconf}
\usepackage{graphicx}
\usepackage{amsmath}
\usepackage{slashed}
\usepackage{epsfig, subfigure}
\def\ga{\mathrel{\raise.3ex\hbox{$>$\kern-.75em\lower1ex\hbox{$\sim$}}}}
\def\la{\mathrel{\raise.3ex\hbox{$<$\kern-.75em\lower1ex\hbox{$\sim$}}}}
\begin{document}
\title{Charged Higgs boson production in single top mode at the LHC }

\author{Renato Guedes$^1$, Stefano Moretti$^2$ and Rui Santos$^{1,3}$}

\address{$^1$Centro de F\'\i sica Te\' orica e Computacional, Faculdade de Ci\^encias, Universidade de Lisboa, Av. Prof. Gama Pinto 2, 1649-003 Lisboa, Portugal}
\address{$^2$ NExT Institute and School of Physics and
Astronomy, University of Southampton Highfield, Southampton SO17 1BJ, UK}
\address{$^3$  Instituto Superior de Engenharia de Lisboa - ISEL, Rua Conselheiro Em\'\i dio Navarro 1, 1959-007 Lisboa, Portugal }

\ead{renato@cii.fc.ul.pt, stefano@phys.soton.ac.uk, rsantos@cii.fc.ul.pt}

\begin{abstract}
The main production mode for a light charged Higgs boson at the LHC is $pp \to t \bar t$, with one
the top-quarks decaying to a charged Higgs and a $b$-quark. However, single top production also gives rise to final states with
charged Higgs bosons. In this work we analyse how the two processes compare at the LHC@14TeV. We will
be working in the framework of the two-Higgs double model, considering both a CP-conserving and a
CP-violating version of the model. We conclude that the single top mode could help to
constrain the parameter space in several versions of the model. We also discuss
the role of other complementary production processes in future searches at the LHC.
\end{abstract}

\section{Introduction}

The discovery of a charged Higgs boson at CERN's Large Hadron Collider would be an unequivocal sign of 
physics beyond the Standard Model (SM). A light charged Higgs is well within reach of the LHC@8TeV in
many Beyond the SM (BSM) models. Searches based on  $pp \to t \bar t \to b W^+ \bar b H^-$ 
are currently being performed by both the ATLAS~\cite{ATLASICHEP} and CMS~\cite{CMSICHEP} collaborations. A large portion of the
parameter space with a light charged Higgs boson (below 150 GeV) has already been excluded 
by the two experiments in models with two Higgs doublets and in particular in the Minimal
Supersymmetric Standard Model (MSSM). However, it is clear that even for a charged Higgs mass
below the top-quark mass, the entire parameter space will not be ruled out when all the 
8 TeV data is finally analysed. Hence, one may ask if by the end of the 13-14 TeV run a light charged 
Higgs boson will be either found or definitely excluded and if so in which models?
The answer to that question is highly dependent on the model being scrutinised. It is expectable that 
most of the parameter space with a charged Higgs mass below the top mass will be probed for the MSSM as well as 
for other multi-Higgs extensions of the SM.
It is clear though, that some multi-Higgs versions will not be probed in their entire parameter space range
even for a light charged Higgs. For those particular scenarios  it is useful to consider all charged Higgs
production process as to maximize the discovery potential. As the single top production cross section is
about one third of the $t \bar t$ one, it could in principle slightly boost the chances of finding or indeed disproving
the existence of a light charged Higgs boson. The purpose of this work  is to show that a slight improvement can
 be obtained by complementing the present search, based on the $t \bar t$ mode, with the search in the single top mode.

\section{Two-Higgs doublet models}

The softly broken $Z_2$ symmetric ($\Phi_1 \rightarrow \Phi_1$,
$\Phi_2 \rightarrow - \Phi_2$) two-Higgs doublet model (2HDM) potential can be written as
\begin{align*}
V(\Phi_1,\Phi_2) =& m^2_1 \Phi^{\dagger}_1\Phi_1+m^2_2
\Phi^{\dagger}_2\Phi_2 + (m^2_{12} \Phi^{\dagger}_1\Phi_2+{\mathrm{h.c.}
}) +\frac{1}{2} \lambda_1 (\Phi^{\dagger}_1\Phi_1)^2 +\frac{1}{2}
\lambda_2 (\Phi^{\dagger}_2\Phi_2)^2\nonumber \\ 
+& \lambda_3
(\Phi^{\dagger}_1\Phi_1)(\Phi^{\dagger}_2\Phi_2) + \lambda_4
(\Phi^{\dagger}_1\Phi_2)(\Phi^{\dagger}_2\Phi_1) + \frac{1}{2}
\lambda_5[(\Phi^{\dagger}_1\Phi_2)^2+{\mathrm{h.c.}}] ~, \label{higgspot}
\end{align*}
where $\Phi_i$, $i=1,2$ are complex SU(2) doublets. Hermiticity of the potential 
forces all parameters except $m_{12}^2$ and $\lambda_5$ to be real. Then, the
nature of $m_{12}^2$ and $\lambda_5$, together with the chosen vacuum configuration,
will determine the CP nature
of the model (see~\cite{Branco:2011iw} for a review). If CP is conserved we end
up with two CP-even Higgs states, $h$ and $H$, and one CP-odd state, $A$. 
If CP is broken, the three spinless neutral states with undefined CP quantum number are usually denoted
by $h_1$,  $h_2$ and $h_3$.
However, as long as the vacuum configuration does not break electric charge, which was shown
to be possible in any 2HDM~\cite{vacstab1}, 
there are in any case two charged Higgs boson states, one charged conjugated to the other. 

In this work we will focus on two specific realisations of 2HDMs, one CP-con\-serv\-ing
and the other explicitly CP-violating~\cite{Ginzburg:2002wt,ElKaffas:2006nt}. 
In the CP-violating version $m_{12}^2$ and $\lambda_5$ are complex and 
the fields' vacuum expectation values (VEVs) are real. Existence of a stationary point
requires $Im (\lambda_5) = v_1 \, v_2 \, Im (m_{12}^2)$. Because the VEVs
are real in both models, a common definition for the rotation angle
in the charged sector $\tan\beta=v_2/v_1$ can be used. 
Extending the $Z_2$ symmetry to the Yukawa sector 
we end up with four independent 2HDMs, the well known~\cite{barger, KY} Type I (only 
$\phi_2$ couples to all fermions), Type II ($\phi_2$ couples to up-type quarks and $\phi_1$ couples to 
down-type quarks and leptons), Type Y or III ($\phi_2$ couples to up-type quarks and 
to leptons and $\phi_1$ couples to down-type quarks)  and Type X or IV ($\phi_2$ couples to all quarks and $\phi_1$ couples to leptons)
(details and couplings can be found in~\cite{Guedes:2012eu}).

We will now very briefly discuss the main experimental constraints affecting the 2HDM parameter space.
The signal in our analysis originates from single top production with the subsequent decay $t \to b H^{\pm} \to b \tau \nu$.
Hence, only the charged Higgs Yukawa couplings are present and therefore 
the only parameters we need to be concerned with are $\tan \beta$ and the charged Higgs
mass.   
Values of $\tan \beta$ smaller than $O (1)$ together with 
a charged Higgs with a mass below $O (100$ GeV$)$ are both disallowed by the constraints~\cite{BB} coming from $R_b$, from $B_q \bar{B_q}$ 
mixing and  from $B\to X_s \gamma$ for all models. Furthermore, data from $B\to X_s \gamma$~\cite{BB2}
 imposes a lower limit of $m_{H^\pm}  \ga 360$ GeV, but only for models Type II and Type Y.
The LEP experiments have set a lower
limit on the mass of the charged Higgs boson of 80 GeV  at 95\% C.L., assuming
$BR(H^+ \to \tau^+ \nu) + BR(H^+ \to c \bar s) +  BR(H^+ \to A W^+) =1$~\cite{LEP}. The bound
is increased to 94 GeV if $BR(H^+ \to \tau^+ \nu)  =1$~\cite{LEP}.
These bounds led us to take $m_{H^\pm} > 90$ GeV and $\tan \beta > 1$ for Type I and X. We will
also present results 
for model Type II, where the bounds on the charged Higgs mass can be evaded due to the presence
of new particles as is the case of the MSSM. The usual theoretical bounds such as the ones coming
from requiring boundness from below of the potential and the ones from requiring perturbative unitarity 
are in this case redundant (same is true for the precision electroweak constraints).

\section{Results and discussion}

As previously discussed, $pp \to t \bar t$ is the best process to search for a light 
charged Higgs boson at the LHC. However, because the single top production cross section
 is about one third of  $\sigma_{pp \to t \bar t}$, it deserves a full investigation regarding its
contribution to the production of charged Higgs bosons. The signal consists 
mainly of a light charged Higgs boson produced via the $t$-channel process 
$pp \to t  \,  j \to H^+ \, \bar  b \, j $ (together with its charge conjugate), 
with the subsequent decay $H^+  \to   \tau^+ \, \nu $,
where $j$ represents a light-quark jet.
In what follows we are considering proton-proton collisions
at a center-of-mass (CM) energy of  $\sqrt{s} = 14$ TeV and a top-quark
mass $m_t = 173$ GeV.  We consider a charged Higgs boson mass interval of 90 to 160 GeV and the analysis
is performed in 10 GeV mass steps.

Maximising the signal-to-background significance
($S/\sqrt B$) makes both the $s$-channel and the $tW$ single-top production
modes negligible - only the $t$-channel process survives the set of cuts imposed.
Signal events were generated with POWHEG~\cite{Alioli:2010xd}
at NLO with the CTEQ6.6M~\cite{Nadolsky:2008zw} PDFs.
The top was then decayed in PYTHIA~\cite{Sjostrand:2006za}. We have considered
only the leptonic decays of the tau-leptons, that is, the signal final state is
$pp \to l \, b \, j \, \slashed{E}$, where $l=e, \mu$ (electrons and muons)
while $\slashed{E}$ means missing (transverse) energy.

The irreducible background, single-top production
with the subsequent decay $t \to b \, W^+$, was also generated with POWHEG.
The main contributions to the reducible background are:
 $t\bar t$ production, $W^\pm ~+$~jets (including not only light quarks and gluons,
but also $c$- and $b$-quarks) 
and the pure QCD background ($jjj$, where $j$ is any jet). 
The $t\bar t$ background was generated
with POWHEG while $W^\pm$ + jets (1, 2 and 3 jets) was
generated with  AlpGen~\cite{Mangano:2002ea}.
Finally, the QCD background was generated with 
CalcHEP~\cite{Pukhov:2004ca} (with CTEQ6ll PDFs). 
The hadronisation was performed with PYTHIA 6.  After hadronisation, DELPHES~\cite{Ovyn:2009tx} 
was used to simulate the detector effects.
For the detector and trigger configurations, we resorted to the ATLAS default definitions.

In order to maximise $S/\sqrt{B}$ we apply the following selection cuts (see~\cite{Guedes:2012eu} for details)

\begin{enumerate}
\item We demand one electron with $p_T > 30$ GeV or a muon with $p_T > 20$ GeV, and $|\eta| < 2.5$ for both leptons. 
\item We veto events with two or more leptons with $p_T > 10$ GeV. This
cut eliminates the leptonic $t \bar t$ background almost completely.

\item We veto events with leptons having $p_T$ above 55 GeV.

\item  Events with missing energy below 50 GeV are excluded. This is a cut that 
dramatically reduces the QCD background.

\item We ask for one and only one $b$-tagged jet with $p_T < 75$ GeV. 
We assume a $b$-tagging efficiency of  0.4 (with $R = 0.7$),
while the misidentification rates for the case of $c$-quark jets we take 0.1 and for lightquark/
gluon jets we adopt 0.01.

\item We reconstruct a "top quark invariant mass" as defined in~\cite{Guedes:2012eu} 
and demand all events to have this invariant mass above 280 GeV. The top quark 
invariant mass distribution for signal and background is shown in figure~\ref{fig:ana}.

\begin{figure}[h]
\centering
\includegraphics[width=5.9in,angle=0]{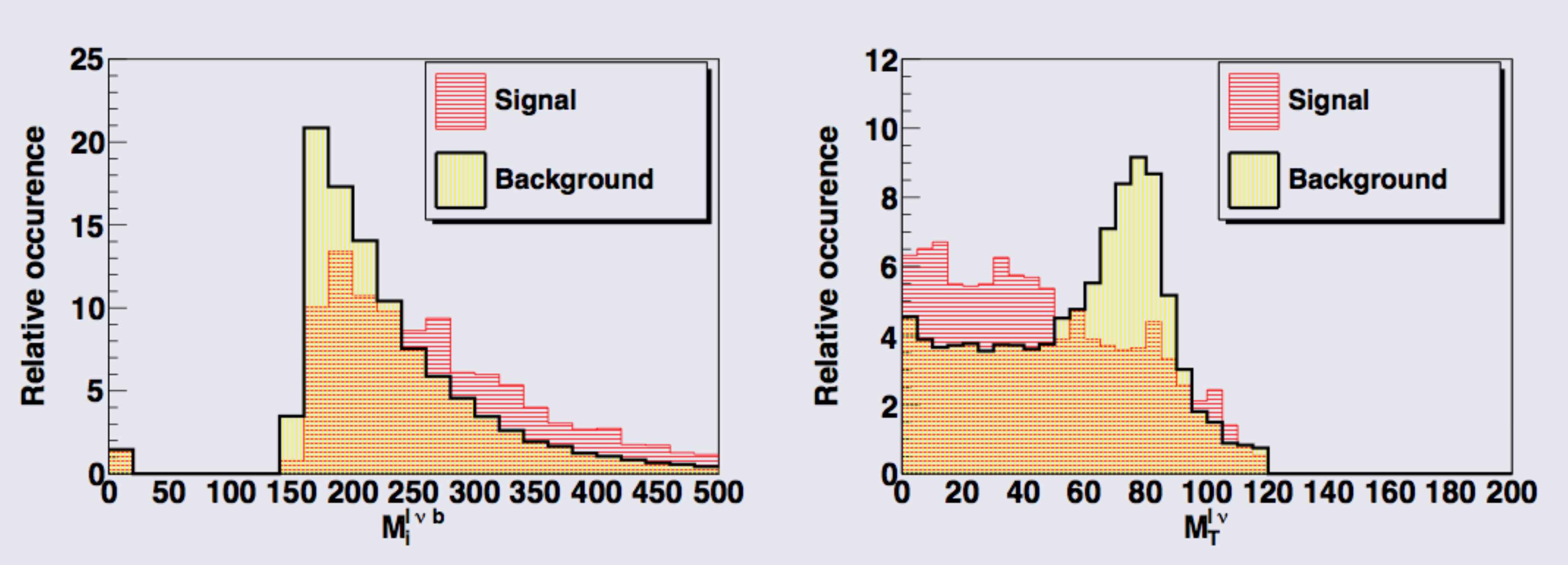}
\caption{Left:  "top quark invariant mass" distribution for signal and background.
Right: leptonic transverse mass distribution for signal and background.}
\label{fig:ana}
\end{figure}

\item We define a leptonic transverse mass~\cite{Guedes:2012eu}, $M_T^{l \nu} $, and we have accepted events with 
$30 \, {\rm GeV} < M_T^{l \nu}  < 60 \, {\rm GeV} $ for charged Higgs masses between 90 and 130 GeV and 
$30 \, {\rm GeV} < M_T^{l \nu}  < 60 \, {\rm GeV} $ or $M_T^{l \nu} > 85 \, {\rm GeV}$ 
for higher values of the charged Higgs mass. The leptonic transverse mass distribution 
for signal and background is shown in figure~\ref{fig:ana}.

\item We have chosen events with one and one jet (non-$b$) only with $p_T>30$ GeV and $|\eta|\leq 4.9$.

\item We veto all events with a jet multiplicity equal to two or above for jets with $p_T>15$ GeV and $|\eta|\leq 4.9$.

\item We only  accept events where jets have a pseudorapidity $|\eta|\geq 2.5$.
\end{enumerate}

Putting all the numbers together we can find  $S/B$ and $S/\sqrt B$
as a function of the charged Higgs mass as presented in table~\ref{tab:signi}.

\begin{table}[h]
\caption{\label{tab:signi}Signal-to-Background ratio ($S/B$) 
and significance ($S/\sqrt B$) as a function of the charged Higgs mass for $\sqrt{s} = 14$ TeV and a luminosity of 1 fb$^{-1}$.
The 
numbers presented for the signal we take BR$(t \to b H^{\pm}) = 100 \%$ and 
BR$(H^- \to \tau^- \nu) = 100 \%$ and all other
Branching Ratios (BRs) have the usual SM values.}
\begin{center}
\lineup
\begin{tabular}{lllll}
\br
$m_H^{\pm}$ (GeV)   & Signal ($S$) & Background ($B$) & $S/B$ $(\%)$ & $S/\sqrt B$\\
\mr
\090 &  \038.6 &  \029.5 & 130.92 & \07.11\\
100 & \040.5 & \029.5 & 137.19 & \07.45\\
110 & \045.6 & \029.8 & 153.00 & \08.35\\
120 & \047.7 & \030.1 & 158.26 & \08.69\\
130 & \042.3 & \032.7 & 129.53 & \07.41\\
140 & 117.1 & \077.9 & 150.25 & 13.26\\
150 & 120.0 & \086.6 & 138.64 & 12.90\\
160 & 109.7 & 100.8 & 108.81 & 10.92\\ 
\br
\end{tabular}
\end{center}
\end{table}

\begin{figure}[h]
\centering
\includegraphics[width=3.0in,angle=0]{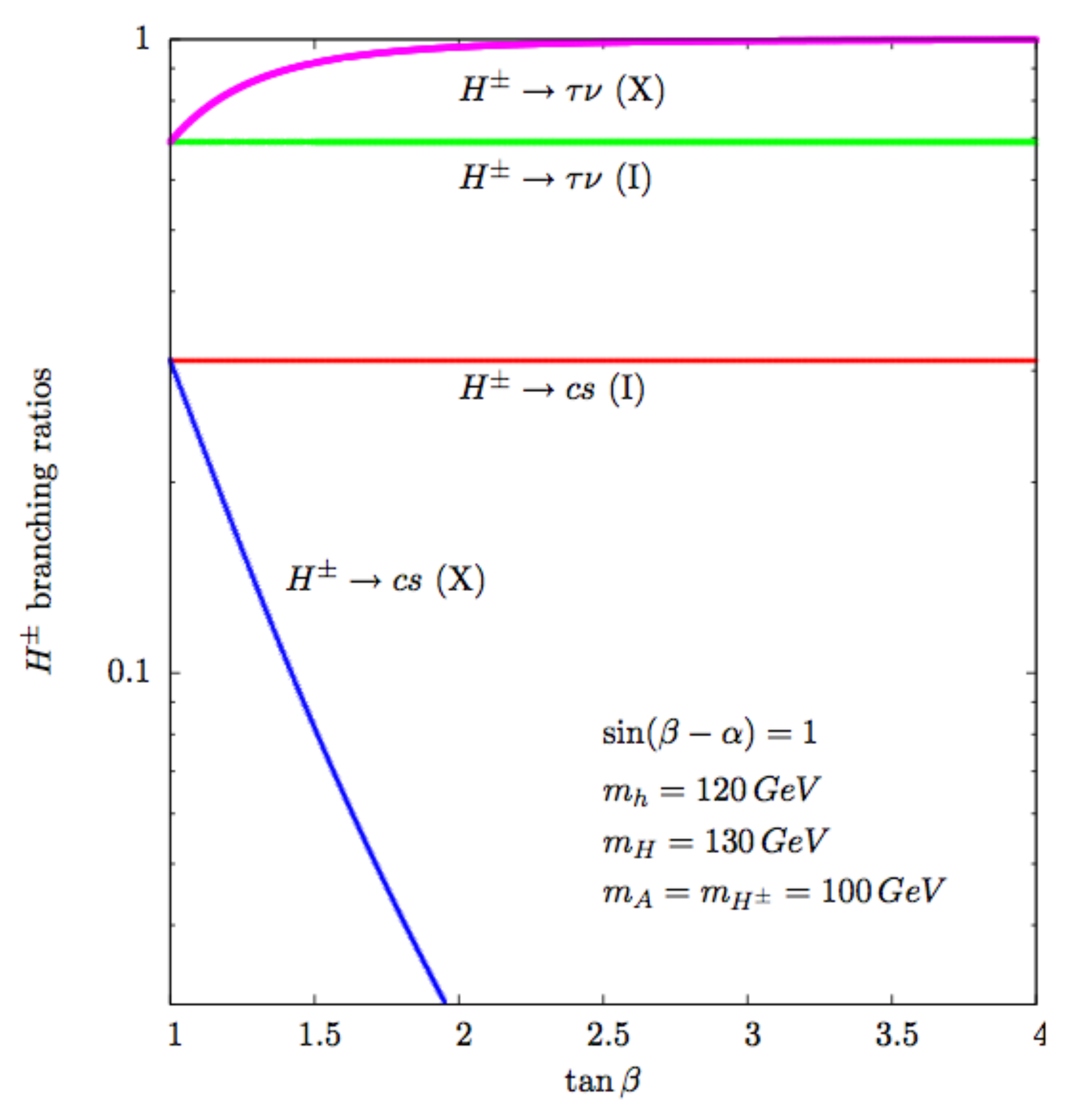}
\hspace{-.1cm}
\includegraphics[width=3.2in,angle=0]{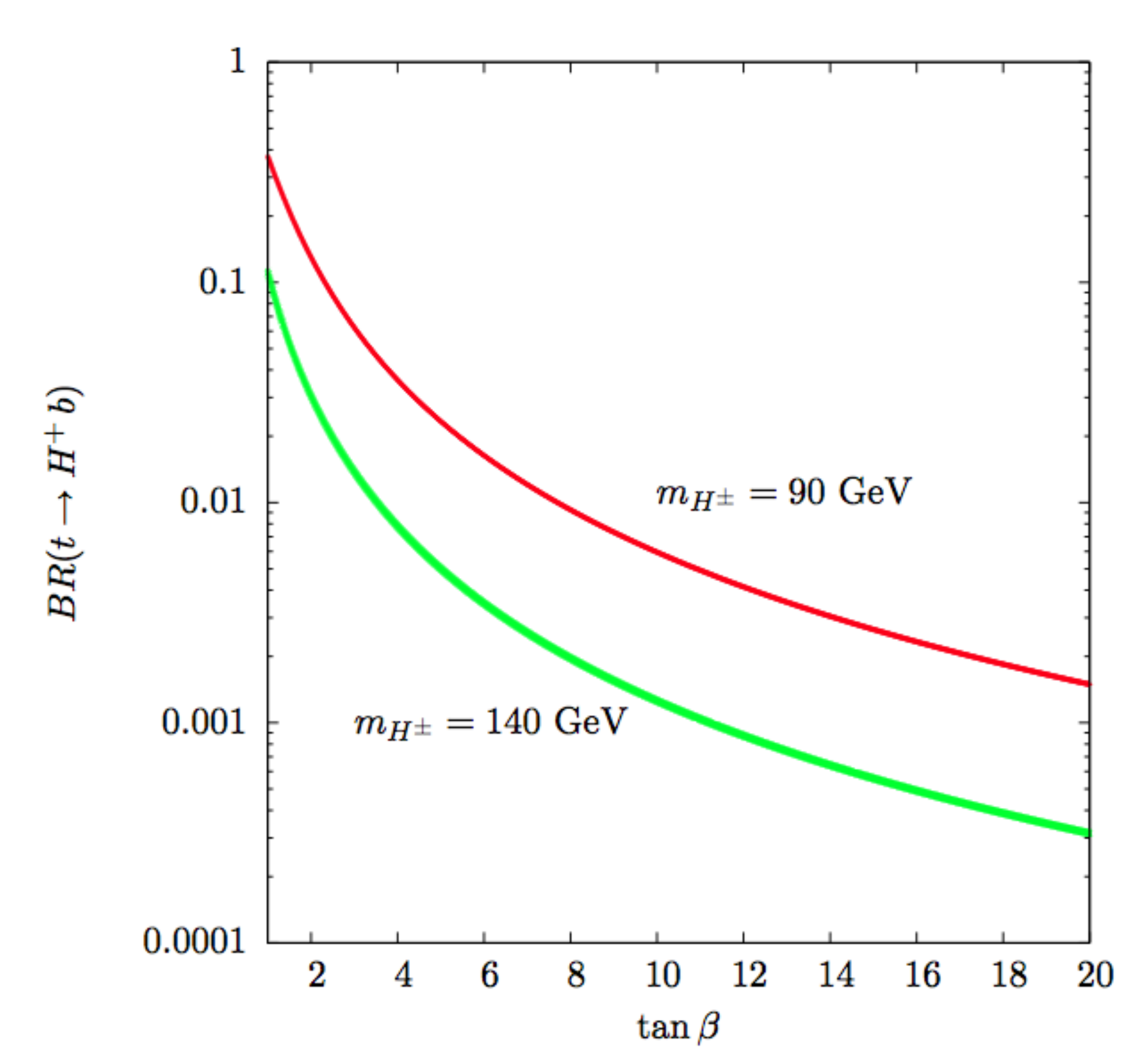}
\caption{Left: charged Higgs BRs for $m_{H^\pm} = 100$ GeV as a function of $\tan \beta$ in models
Type I and X. Right: BR$(t \to H^+b)$ as a function of $\tan \beta$ for two values of the charged Higgs
boson mass.}
\label{fig:BR}
\end{figure}

The results can be presented in a model independent manner as done in~\cite{Guedes:2012eu}
and from them exclusion plots can be derived for the different 2HDMs. Before proceeding we
present in the left panel of figure~\ref{fig:BR} the charged Higgs BRs for $m_{H^\pm} = 100$ GeV
 as a function $\tan \beta$ in models Type I and X. Clearly $H^+ \to \tau^+ \nu$ is the dominant
decay mode in both models. As the charged Higgs boson width
depends only on $\tan \beta$ and on the charged Higgs mass, the plot is representative of 
all values of $m_{H^\pm}$ provided that decays to other neutral scalars is forbidden.
In the right panel of figure~\ref{fig:BR} we show the BR$(t \to H^+b)$ as a function of $\tan \beta$
for two values of the charged Higgs boson mass. Contrary to the case of the MSSM and MSSM-like versions of a Type II
2HDM, this BR falls very rapidly with $\tan \beta$ and even more so as the charged Higgs boson
mass approaches the top-quark mass.

\begin{figure}[h]
\centering
\includegraphics[width=6.3in,angle=0]{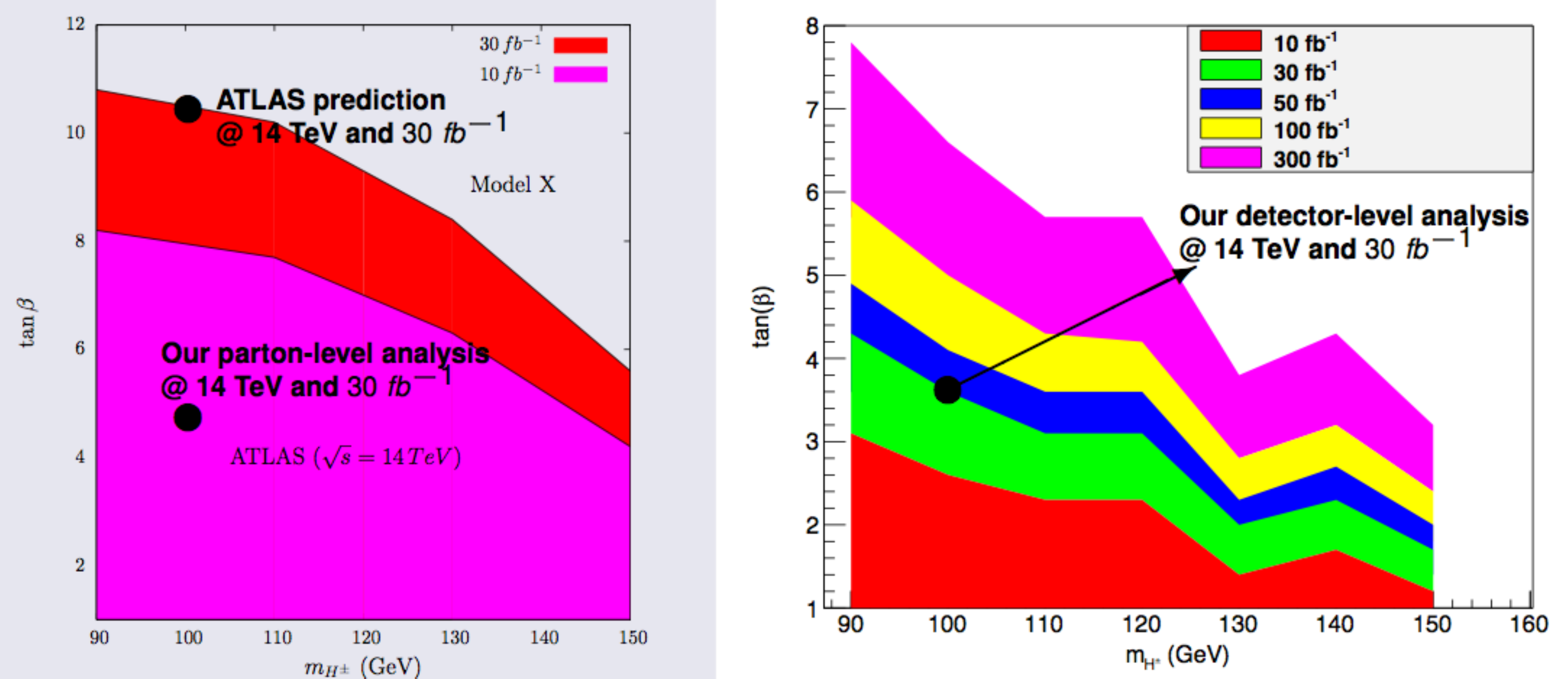}
\caption{Left: Excluded region at the 95\% CL in the ($\tan \beta$, $m_{H^\pm}$)  plane
for the Type X 2HDM using the ATLAS predictions
for 10 fb$^{-1}$ and 30 fb$^{-1}$ of total integrated luminosity for the  LHC@14 TeV~\cite{Aad:2009wy}.
Also shown is our parton level prediction for 30 fb$^{-1}$~\cite{Aoki:2011wd}.
Right: excluded region in the  ($\tan \beta$, $m_{H^\pm}$)  plane
for Type X at the 95\% CL assuming the LHC@14 TeV and for several luminosity sets.}
\label{fig:MX}
\end{figure}

\begin{figure}[h]
\centering
\includegraphics[width=6.3in,angle=0]{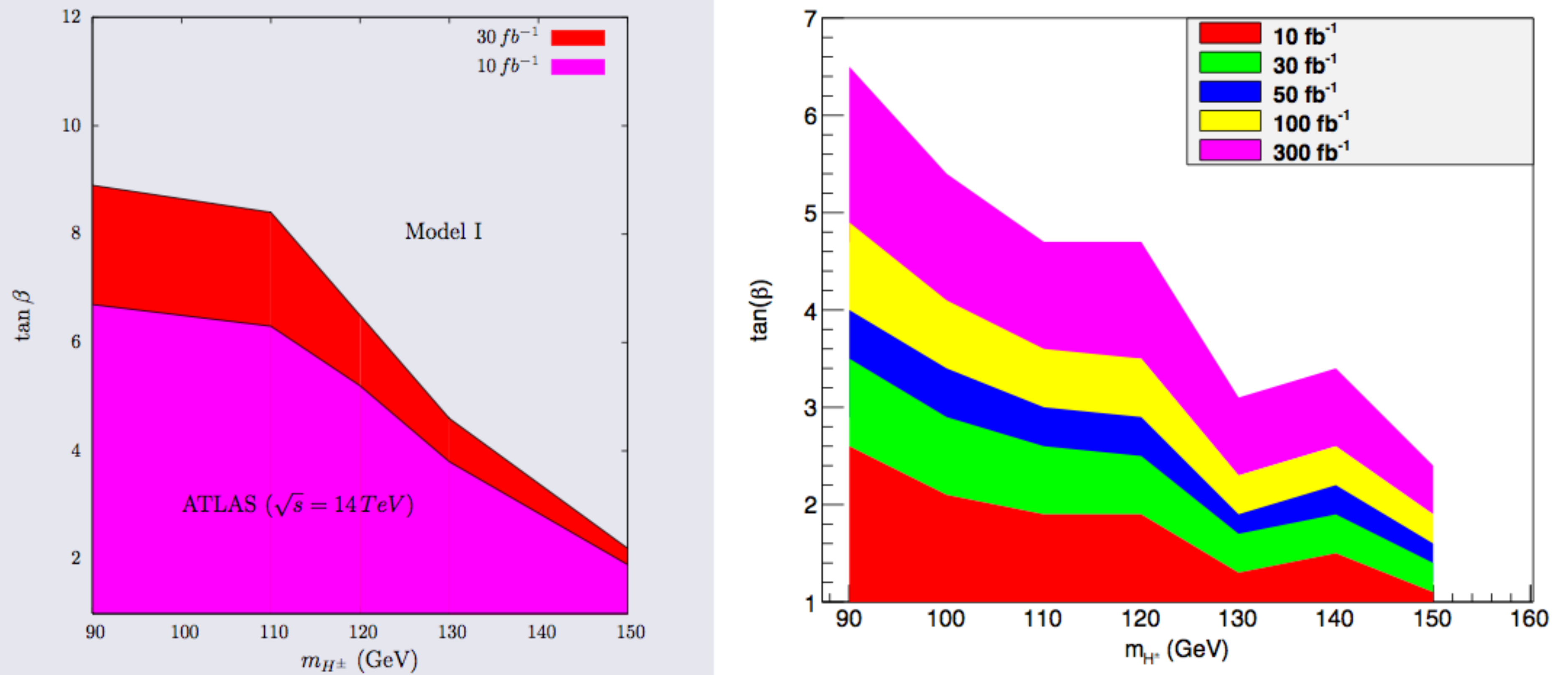}
\caption{Left: excluded region at the 95\% CL in the ($\tan \beta$, $m_{H^\pm}$)  plane
for the Type I 2HDM model using the ATLAS predictions
for 10 fb$^{-1}$ and 30 fb$^{-1}$ of total integrated luminosity for the  LHC@14 TeV~\cite{Aad:2009wy}.
Right: excluded region in the  ($\tan \beta$, $m_{H^\pm}$)  plane
for Type I at the 95\% CL assuming the LHC@14 TeV and for several luminosity sets.}
\label{fig:MI}
\end{figure}

Using this information we can now draw exclusion plots for the different 2HDM types.
In the left panel of 
figure~\ref{fig:MX} we present the excluded region at the
95\% CL in the ($\tan \beta$, $m_{H^\pm}$)  plane
for the Type X 2HDM model using the ATLAS predictions
for 10 fb$^{-1}$ and 30 fb$^{-1}$ of total integrated luminosity for the LHC@14 TeV~\cite{Aad:2009wy}.
In the same figure we have drawn a point that corresponds to our parton 
level prediction for 30 fb$^{-1}$ and $m_{H^\pm} = 100$ GeV presented in~\cite{Aoki:2011wd}.
In the right panel of the same figure we show the excluded region in the  ($\tan \beta$, $m_{H^\pm}$)  plane
for Type X at the 95\% CL assuming the LHC@14 TeV and for several luminosity sets.
The 100 GeV mass point at 30 fb$^{-1}$ is also shown for a better comparison both with the ATLAS
prediction and with our previous parton level study~\cite{Aoki:2011wd}.
In the left panel of figure~\ref{fig:MI} we now present the excluded region 
for the Type I model again using the ATLAS predictions
for 10 fb$^{-1}$ and 30 fb$^{-1}$ of total integrated luminosity for the  LHC@14 TeV~\cite{Aad:2009wy}.
In the right panel we present our final detector level results for Type I 
at the 95\% CL assuming the LHC@14 TeV and for several luminosity sets.
We can conclude from the plots that as expected the results show similar trends to the ones obtained for
$t \bar t$ production. We started with a cross section that
is about three times smaller than the $t \bar t$ one and ended up
with a result that is 2 to 3 times worse than the prediction
presented by ATLAS~\cite{Aad:2009wy}.

It is expectable that both ATLAS and CMS
will improve the results on the single top mode presented here, tightening
the constraints on the $(m_{H^\pm}, \tan \beta)$ plane.
One may now ask what are the chances to probe the entire $(m_{H^\pm}, \tan \beta)$ plane
by the end of the 14 TeV run. 
In view of the results for 7 TeV~\cite{ATLASICHEP, CMSICHEP}, one expects a type II light charged Higgs to be excluded
by then. 
However, there are models
where the Yukawa couplings always decrease with $\tan \beta$ as is the case
of models I and X. For those models, we know that $pp \to t \bar t$ will provide
the strongest constraint on the $(m_{H^\pm}, \tan \beta)$ plane, and that the single
top mode is bound to contribute even if only with a slight improvement.
Are there any other processes that could help to probe the large $\tan \beta$ region?

There is another Yukawa process, $cs  \to H^{\pm}  (+j)$~\cite{cs, Aoki:2011wd},  that could in principle help to probe the 
above mentioned region. It was however shown to be negligible for large $\tan \beta$.  
The remaining possibility~\cite{Aoki:2011wd} is to look for processes that
either do not depend on $\tan \beta$, or even better, that grow
with $\tan \beta$. There are terms both in $gg \to H^+W^-$ 
and in vector boson fusion ($pp \to jj H^+H^-$ where $j$ is a light quark jet) that are independent of $\tan \beta$.
Furthermore, for the CP-conserving potential, there is a term in $gg \to H^+H^-$ that 
has the form 
\begin{equation}
\sigma_{pp \to H^+ H^-} \propto \sin (2 \alpha) \, \tan \beta (m_H^2 -  M^2)
\end{equation}
where $\alpha$ is the rotation angle in the CP-even sector, $m_H$ is the heavier CP-even scalar mass
and $M^2 = m_{12}^2/(\sin \beta \, \cos \beta)$. Hence, there are regions of the 2HDM
parameter space that can be probed for larger values of $\tan \beta$. However, the bounds
will no longer be for a two parameter space but instead for a multi-dimension space 
with all 2HDM parameters playing a role. Further, values of the cross section
that could lead to meaningful significances are only obtained for resonant production. Therefore,
only a small portion of the multi-dimensional space can be probed for large $\tan \beta$ (see~~\cite{Aoki:2011wd} for details). 

A final comment about theoretical bounds. Assuming that the Higgs boson was
discovered with a mass of 125 GeV, it was recently shown in~\cite{Maria}
that  for the particular case of an exact CP-conserving $Z_2$ symmetric model
$\tan \beta < 6$. Therefore, that particular model will probably see a light charged
Higgs ruled out when all the 8 TeV data is analysed.

\ack{ 
SM is financed in part through the NExT Institute. 
The work of RG and RS is supported in part by the Portuguese
\textit{Funda\c{c}\~{a}o para a Ci\^{e}ncia e a Tecnologia} (FCT)
under contracts PTDC/FIS/117951/2010 and PEst-OE/FIS/UI0618/2011.
RG is also supported by a FCT Grant SFRH/BPD/47348/2008.
RS is also partially supported by an FP7 Reintegration Grant, number PERG08-GA-2010-277025.}

\end{document}